\documentclass[reprint,amsmath,amssymb,aps]{revtex4-2}
\usepackage{graphicx}
\usepackage{dcolumn}
\usepackage{bm}
\usepackage[utf8]{inputenc}
\usepackage{footmisc}
\graphicspath{ {./images/} }
\usepackage{natbib}
\usepackage{hyperref}
\usepackage{url}
\hypersetup{
   colorlinks=true,
   linkcolor=blue,
   citecolor=blue,
   urlcolor=blue,
   }
\usepackage[mathlines]{lineno}
\begin{document}
\preprint{APS/123-QED}
\title{Tuneable vertical hysteresis loop shift in Ni$_{80}$Fe$_{20}/$SrRuO$_3$ heterostructures}

\author{Manisha Bansal$^{1}$}

\author{Samir Kumar Giri$^{2}$}
\author{Weiwei Li$^{2}$}
\author{Judith L. MacManus-Driscoll$^{2}$}
\author{Tuhin Maity$^{1,2}$}
 \email{tuhin@iisertvm.ac.in}

\affiliation{$^{1}$School of Physics, Indian Institute of Science Education and Research Thiruvananthapuram, Thiruvananthapuram, Kerala 695551, India\\
$^2$Department of Materials Science and Metallurgy, University of Cambridge, 27 Charles Babbage Road, Cambridge CB3 0FS, United Kingdom}

\date{\today}

\begin{abstract}
A novel vertical hysteresis loop shift is observed for a bilayer thin film of Ni$_{80}$Fe$_{20}$/SrRuO$_3$ (1:2 thickness ratio) on SrTiO$_3$ substrate, after field cooled through the Curie temperature $\sim$125 K of SrRuO$_3$, whereas no shift is found for zero field cooled measurement. The vertical shift is noticed below $\sim$125 K which increases with decreasing temperature and reaches a maximum of 33\% at 2 K. The vertical shift is modelled closely using micromagnetic simulations. It is shown that the amount of vertical shift depends on the competition between the relative thicknesses and the anisotropies of the two films governed by Spring Thickness law and Imperial Law, respectively. We propose a generalized model of vertical shift which is applicable for other material systems, and hence could be informative for use of vertical shift in future spintronic devices.
\end{abstract}
\maketitle
\section{Introduction}
Exchange bias (EB) has garnered huge amount of interest due to its tremendous success in several breakthrough applications in spintronics\cite{childress2001irmn,allibe2012room}. Decades of research have been carried out in search for fascinating EB related phenomena in several material systems. In conventional EB system, the magnetic hysteresis loop ($MH$) shifts in the horizontal or field ($H$) axis\cite{nogues1999exchange,nogues2005exchange} due to the interfacial coupling at ferromagnetic (FM)/antiferromagnetic (AFM) interface. Researchers have been curious to find the $MH$ shift along the magnetization ($M$) axis, i.e., a vertical bias (VB) in continuation of the conventional EB for years. This shift allows an extra degree of freedom to be harnessed, potentially giving additional functionality in future extraordinary spintronic devices, e.g., multi-state magnetic recording technology, and also perhaps greater robustness\cite{zheng2016tunable,tian2020giant,avci2017multi}. 
The vertical shift like behavior has been observed in few systems including magnetic heterostructures, multiferroic ceramics, core-shell magnetic nanoparticles, manganite based superlattices, etc\cite{rana2014positive,buchner2019transition,singamaneni2015complete,zhou2020orbital,zheng2004giant,tian2010spin}. Some of these reports have indicated that the interfacial spin glass state arising from the uncompensated spins may account for such shift\cite{zheng2004giant,tian2010spin}. On the other hand, other researches have indicated that vertical shift could result from variations in the domain structure leading to uncompensated spins or pinned magnetic moments at the interface as found in some FM/AFM systems\cite{rana2014positive,singamaneni2015complete,ohldag2003correlation,fitzsimmons2007pinned,rana2014controlling}. However, the vertical shift has been reported even with a compensated G-type AFM\cite{yuan2013giant}. So the actual reason for the elusive VB is debatable due to the discrepancies so far. No research has come up with the concrete results with a quantitative model of tuneable VB and its explanation. Indeed, most of the research to date has reported only an insignificant and non-tuneable shift, which is not so promising for applications. Although observations of VB like behavior in some systems have intrigued the scientific community, a quantitative modelling and an in-depth understanding for tuneable VB is still outstanding.

In this report, we show a clear VB ($\sim$33\%) in a Ni$_{80}$Fe$_{20}$/SrRuO$_3$ (SRO) bi-layer thin film at 2 K after cooling under ±1 T bias field and $MH$ loop measured with ±0.5 T field range. The zero field cooled (ZFC) measurement does not show any VB similar to the conventional horizontal EB behavior. Temperature ($T$) dependent VB measurements are done to show the evolution of VB. Micromagnetic simulations are also undertaken which are found in good agreement with experimental results. We show that the VB strongly depends on the thickness, anisotropy, temperature, etc. By using both experimental and simulation results, we derive a generalized model of VB to understand its origin and to explore VB in different material systems for its potential applications in spintronic devices.

\section{Experimental methods}
13-unit cells of epitaxial SRO film ($\sim$5 nm) were deposited on a SrTiO$_3$ (STO) substrate by advanced pulsed laser deposition (PLD) technique. The thickness was monitored by Reflection high-energy electron diffraction (RHEED) in PLD. On top of that, a 2 nm Ni$_{80}$Fe$_{20}$ layer was deposited by DC/RF sputtering. The detailed magnetic measurements were carried out in a Quantum Design MPMS3 magnetometer. When it was required, the sample was demagnetized at RT by a proper demagnetization protocol and the magnet was reset before the measurement to ensure that there was no trapped flux present in the sample or the superconducting coils of the magnetometer\cite{maity2013superspin}. ZFC and positive/negative field cooled (FC) $MH$ measurements were carried out similar to conventional EB measurements with both bias fields of ±1 T and ±7 T to observe VB, if any. For ZFC $MH$ measurement, the sample was cooled down from 350 K to 2 K under no bias field and then the loop was swept in the loop tracing field range of 0 T→ +0.5 T→ -0.5 T→ +0.5 T while for +FC and -FC $MH$ measurement, the sample was cooled from 350 K to 2 K under bias field and the loop was swept in the loop tracing field range of +0.5 T→ -0.5 T→ +0.5 T (under +1 T bias field) and -0.5 T→ +0.5 T→ -0.5 T (under -1 T bias field), respectively. Moreover, the FC $MH$ was measured at various temperatures to investigate $T$ dependent VB study. The amount of the vertical shift of $MH$, $M_{\textrm{VB}}$ was calculated as ($M_{\textrm{S+}}$ + $M_{\textrm{S-}}$)/2, where $M_{\textrm{S+}}$ ($M_{\textrm{S-}}$) corresponds to the positive (negative) saturation magnetization ($M_{\textrm{S}}$) value (signs are included). The magnetization \emph{vs.} temperature ($MT$) measurements for ZFC, FC and remanence (REM) were carried out to identify the phase transition of the film. For ZFC $MT$ measurement, the sample was first demagnetized at room temperature, cooled down from 350 K to 2 K with zero bias field. A field of 1 mT was applied and the $M$ was measured as a function of $T$ while ramping from 2 K up to 350 K. Then, similar measurement was done immediately for FC $MT$ measurement while cooling from 350 K to 2 K under 1 mT bias field. Then the applied field was removed and magnetization was measured while heating from 2 K to 350 K to get REM $MT$. 

\section{Simulation methods}
In order to verify the experimental results, Object Oriented Micromagnetic Framework (OOMMF) simulations were performed\cite{donahue1999oommf}. A bilayer system of 20 nm SRO film and 10 nm Ni$_{80}$Fe$_{20}$ was simulated with the thickness ratio ($t_{\textrm{R}}$) 2:1 almost tallying with the $t_{\textrm{R}}$ of the two films taken experimentally. The total area of the atlas was taken as 320 nm x 320 nm. Only in plane 75\% of the middle portion of the atlas (Fig. S2)  was considered magnetic to avoid edge effects\cite{maity2012ordered}. Rest of the area of the atlas was non-magnetic. The cell dimensions were taken as 10 nm x 10 nm x 5 nm along x, y and z directions, respectively to include exchange interaction effect for Ni$_{80}$Fe$_{20}$ and SRO. These cell dimensions were kept constant all through the simulation process to maintain the consistency of the results. The typical simulation parameters were $M_{\textrm{S}}$ of 600 kA/m (Ni$_{80}$Fe$_{20}$) and 350 kA/m (SRO), anisotropy constant ($K_{\textrm{1}}$) of 0 kJ/m$^3$ (Ni$_{80}$Fe$_{20}$)\cite{boardman2005computer} and 500 kJ/m$^3$ (SRO)\cite{ziese2010inverted} and exchange stiffness constant ($A$) of 13 pJ/m (Ni$_{80}$Fe$_{20}$)\cite{boardman2005computer} and 1.8 pJ/m (SRO)\cite{ziese2010inverted}. In our case, a pulse of a very high field was applied initially for a short time to replicate the FC biasing-like experiment. Then, $MH$ were simulated in the range of ±0.9 T and ±7 T in a sequence to understand the field dependence on VB. $M_{\textrm{VB}}$ was calculated with the remanence magnetization ($M_{\textrm{R}}$) values, i.e., ($M_{\textrm{R+}}$ + $M_{\textrm{R-}}$)/2, where $M_{\textrm{R+}}$ ($M_{\textrm{R-}}$) is the positive (negative) $M_{\textrm{R}}$ (signs are included) correspond to the moment in descending (ascending) branch of the $MH$ at zero magnetic field. Further, due to the linear change of $H_{\textrm{C}}$ with changing $t_{\textrm{R}}$ and $K_{\textrm{1}}$ (Fig. S3), $M_{\textrm{VB}}$ was divided by $H_{\textrm{C}}$ to normalize its effect. Thickness ($t$) and $K_{\textrm{1}}$ of the films were also varied to estimate their effect on $M_{\textrm{VB}}$ manifesting a generalized ground. Since, the total $M_{\textrm{S}}$ can be manipulated by the change of thickness, we have not considered its effect as both of them are strongly correlated. Also, the varying $A$ had no noteworthy effect over the vertical shift of the $MH$ so we kept it constant throughout. We also changed the anisotropy direction, and found $MH$ for (113) direction to be most comparable with the experimental results ($MH$ plots for different anisotropy directions (Fig. S7) are provided in SI).

\section{Experimental Results}
Fig. 1 shows the $MH$ curves for ZFC (grey) and ±FC (±1 T) (blue and red) measurements for a bilayer system of Ni$_{80}$Fe$_{20}$/SRO films on STO (110) substrate at 2 K. The field range of $MH$ were taken to be ±0.5 T to confirm the saturation of Ni$_{80}$Fe$_{20}$. The grey curve clearly shows no VB for ZFC measurement whereas a positive (negative) VB is observed for positive (negative) bias field of ±1 T FC. The ±FC loops shift $\sim$±33\% at 2 K compared to the ZFC loop. The significant decrease of the coercivity ($H_{\textrm{C}}$) from $\sim$4.7 mT in ZFC to $\sim$2.8 mT in ±FC is also observed (Fig. 1(b)). Besides, a horizontal EB of magnitude $\sim$1.7 mT is observed for VB curves from $(H_{\textrm{C+}} + H_{\textrm{C-}})/2$, where $H_{\textrm{C+}}$ and $H_{\textrm{C-}}$ are the coercivities (signs are included) in the ascending and descending branches, respectively. Further, similar $MH$ measurements were done using a high field range of ±7 T for FC (7 T bias) and ZFC (olive green curve in Fig. 2(a)) clearly depicting no VB whereas shifts are observed for ±0.5 T field range (blue and red curve), also shown in Fig. 1. 
\begin{figure}[ht!]
	\centering
	\includegraphics[width=8.5cm]{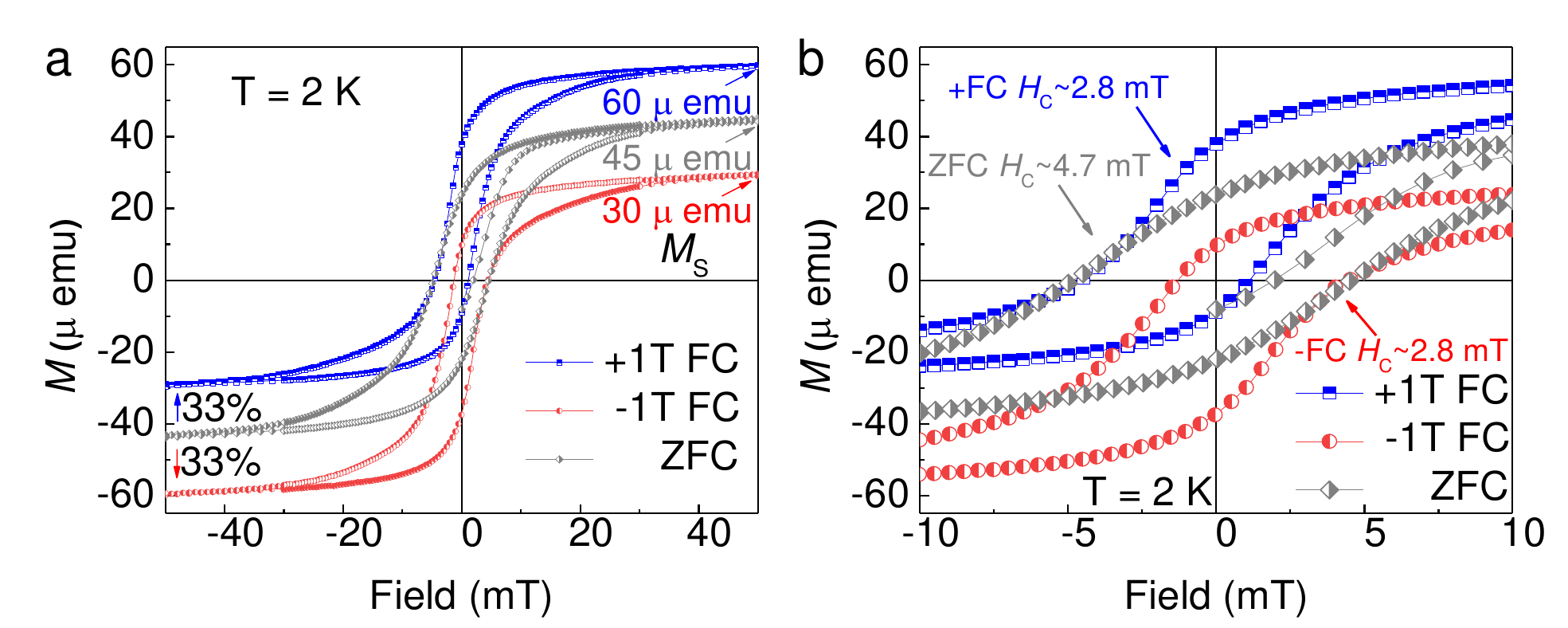}
	\vspace{-1em}
	\caption{(a) ZFC (grey) - ±FC (blue and red) measurements for a bilayer system of Ni$_{80}$Fe$_{20}$/SRO films on STO substrate at 2 K. (b) Zoomed-in version of the main figure illustrates the $H_{\textrm{C}}$ change.}
	\vspace{-0.7em}
\end{figure}

\begin{figure}[ht!]
	\centering
	\includegraphics[width=8.5cm]{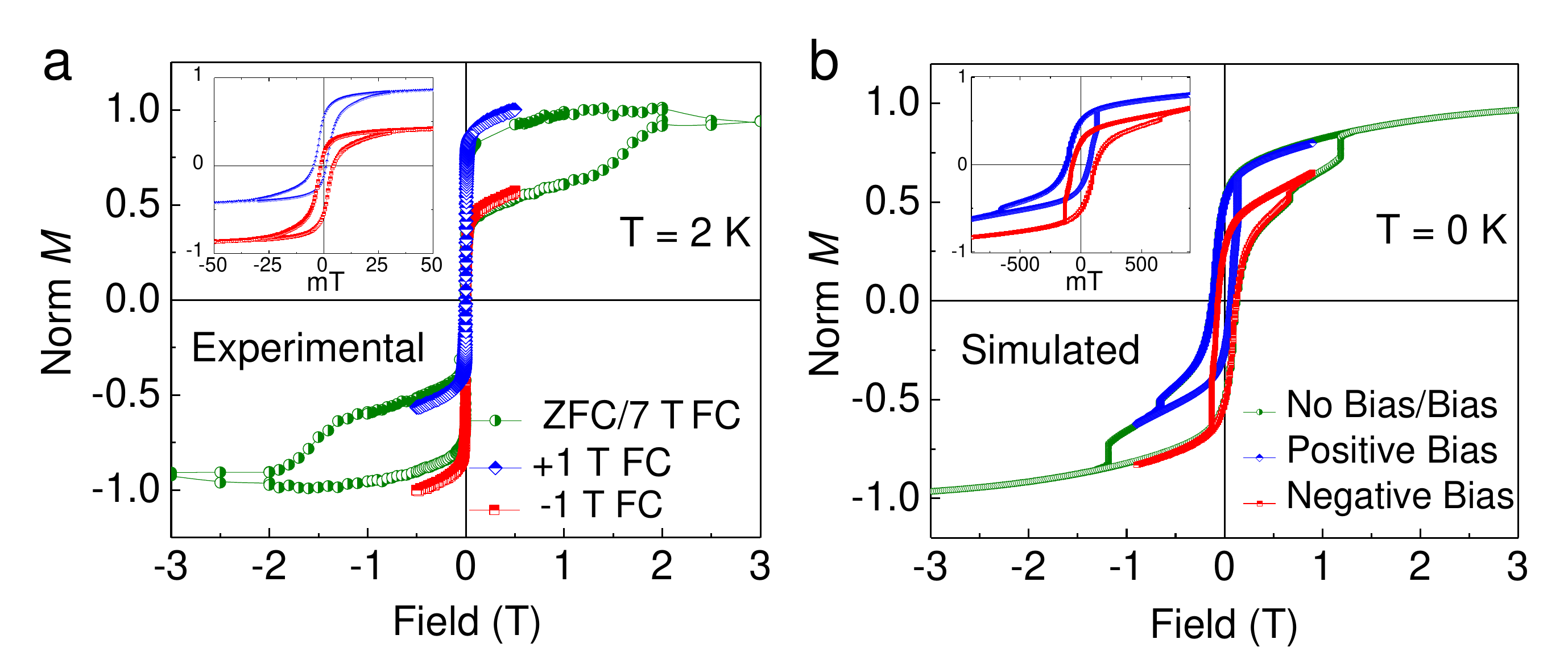}
    \vspace{-1em}
	\caption{(a) Experimental result of VB for Ni$_{80}$Fe$_{20}$/SRO heterostructure measured at 2 K with 7 T bias field and zero bias field/±7 T field range $MH$ (green) and ±1 T bias field/±0.5 T field range $MH$ (blue and red). (b) Simulation result of VB similar to the experimental results. Green curve for ±7 T loop tracing field range and blue and red shifted curves for high pulse and ±0.9 T field range. (Insets) Zoomed positive and negative shifts are explicitly shown.}
	\vspace{-0.7em}
\end{figure}
\begin{figure*}[ht!]
	\centering
	\includegraphics[width=15cm]{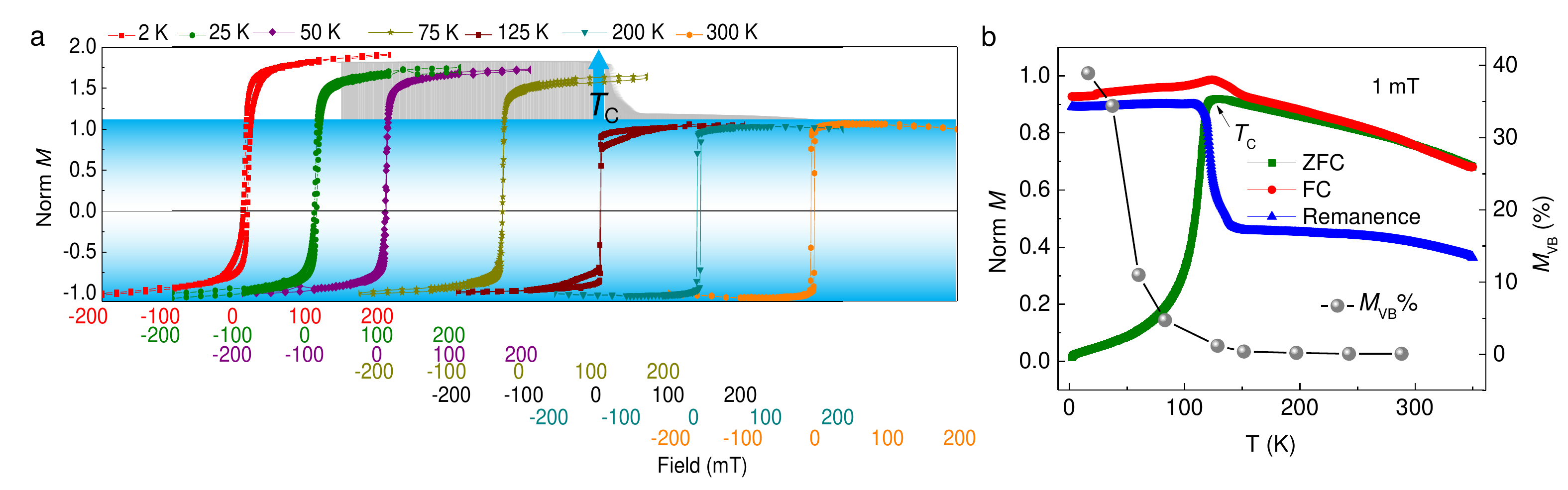}
	\vspace{-1em}
	\caption{(a) Trend of asymmetricity of the hysteresis loops at different $T$. Above $\sim$125 K, no shift is seen. (b) Variation of $M_{\textrm{VB}}$ (\%) with $T$ in black circles. FC (red), ZFC (green) and REM (blue) $MT$ measurements show the phase transition point at the $T_{\textrm{C}}$ of 125 K. The blue curve refers to the REM \emph{vs.} $T$ plot extracted from the experiments similar to the $M_{\textrm{VB}}$ (\%) \emph{vs.} $T$ curve.}
	\vspace{-1em}
\end{figure*}
We carried out temperature-dependent measurements to explore the evolution of VB since the competition among the Zeeman energy ($E_{\textrm{Z}}$), anisotropy energy ($E_{\textrm{A}}$) and thermal energy ($E_{\textrm{T}}$) should be considered in accordance with the Stoner-Wohlfarth model\cite{buchner2019transition}. From the $MH$ curves recorded at different $T$, a decreasing trend of $M_{\textrm{VB}}$ from low $T$ (2 K) to $\sim$125 K is seen and it becomes zero at temperatures higher than $\sim$125 K (Fig. 3(a)). The $M_{\textrm{VB}}$ with different $T$ has been plotted (black circles) in Fig. 3(b).
The $MT$ (ZFC-FC-REM) measurement in Fig. 3(b) shows the Curie temperature ($T_{\textrm{C}}$) of SRO to be $T_{\textrm{C}}$ $\sim$ 125 K\cite{ziese2010structural,zhang2001thickness}, the exact temperature below which VB is observed (explicitly illustrated in Fig. 3(a)). Essentially, SRO becomes FM from paramagnetic (PM) below $T_{\textrm{C}}$ $\sim$ 125 K while cooling down from 350 K to 2 K whereas Ni$_{80}$Fe$_{20}$ is FM throughout (because of its high $T_{\textrm{C}}$ : 500-800 K)\cite{yin2006magnetocrystalline,ekholm2010influence} confirmed from its steady positive REM above $T_{\textrm{C}}$ $\sim$ 125 K of SRO. Note that the trend followed by the the REM \emph{vs.} $T$ curve (blue) is similar to the calculated $M_{\textrm{VB}}$ \emph{vs.} $T$ (black circles) implying the close relation of REM of SRO with the $M_{\textrm{VB}}$. Both the curves show a steep rise once they reach the $T_{\textrm{C}}$ of SRO ($T_{\textrm{C}}$  $\sim$ 125 K).

\section{Simulation Results}
From OOMMF simulations, we found that the loop swept in the range of ±0.9 T field, achieves VB (blue and red $MH$ in Fig. 2(b)). The positive (negative) shift of the $MH$ loop was observed for positive (negative) bias field, similar to the experimental shifts obtained (Fig. 2(a)). On the contrary, the hysteresis loop swept in the field range ±7 T shows no VB (olive green $MH$ Fig. 2(b)) exactly homogenous to the experimental curves (Fig. 2(a)) achieved.

From the simple Meiklejohn-Bean model for the EB system which states the reliance of horizontal shift over the $t$ and $M$ of the films in the form of an energy expression\cite{binek2001exchange}, we can intuitively extend the dependence of these parameters along with the anisotropy over $M_{\textrm{VB}}$. A variation of these parameters is believed to achieve a tuneable VB, consequently giving a valuable insight into the mechanism. We have studied the effect of anisotropy and thickness on $M_{\textrm{VB}}$.
The effect of $K_{\textrm{1}}$ on $M_{\textrm{VB}}$ is shown in Fig. 4(a) where $K_{\textrm{1}}$ is varied from very low value of $200\; kJ/m^3$ to very high value of $2000\; kJ/m^3$. This includes well-known hard materials (e.g., Co, MnAl, FePd, etc.) with their $K_{\textrm{1}}$ marked in Fig. 4(a) to compare their $M_{\textrm{VB}}$ shift. $MH$ plots for different $K_{\textrm{1}}$ is given in SI Fig. S5. It is found that large values of $K_{\textrm{1}}$, i.e., $K_{\textrm{1}} > 600\; kJ/m^3$ do not lead to further increase in $M_{\textrm{VB}}$. At a certain point, the proportion of the two layers reaches the maximum compatibility ratio after which VB becomes almost constant. Hence, the data points up to $K_{\textrm{1}} \sim 600\; kJ/m^3$ are considered for fitting to estimate the shift change with respect to anisotropy. These data points have been successfully fitted (solid green curve) by the Imperial Law (Eq. \eqref{eq1}) which states an even power dependence of $M_{\textrm{VB}}$ on $K_{\textrm{1}}$\cite{zimmermann2007micromagnetic}.
\begin{equation}
      M_{\textrm{VB}} = A+B{K_{\textrm{1}}}^2+CK_{\textrm{1}}^4    \label{eq1}         
\end{equation}
where A, B, and C are the fitting parameters (values are given in SI Table S1).

In addition, we varied $t_{\textrm{R}}$ (hard/soft) of the two layers to attain the thickness dependence on $M_{\textrm{VB}}$ by simulations (the $MH$ plots are given in SI Fig. S4). The trend is depicted in the inset of Fig. 4(b). In the same figure, we also did a best fit (solid blue curve) of the thickness ratio data points (blue spheres) with the Spring Thickness Law (Eq. \eqref{eq2}) which is analogous to the $t$ dependence of the horizontal EB version\cite{binek2001exchange}. 
\begin{equation}
      M_{\textrm{VB}} =  D+E/t_{\textrm{R}}+F/({t_{\textrm{R}}}^2)   \label{eq2}       
\end{equation}
where D, E and F are the involved fitting parameters (values are given in SI Table S1). 
The curve implicates that the maximum $M_{\textrm{VB}}$ is obtained for the 2:1 ratio for the bilayer system of SRO and Ni$_{80}$Fe$_{20}$ (close to experimental ratio). The shift increases rapidly to this $t_{\textrm{R}}$ while it goes down and then saturates.

\begin{figure}[ht!]
	\centering
	\includegraphics[width=8.5cm]{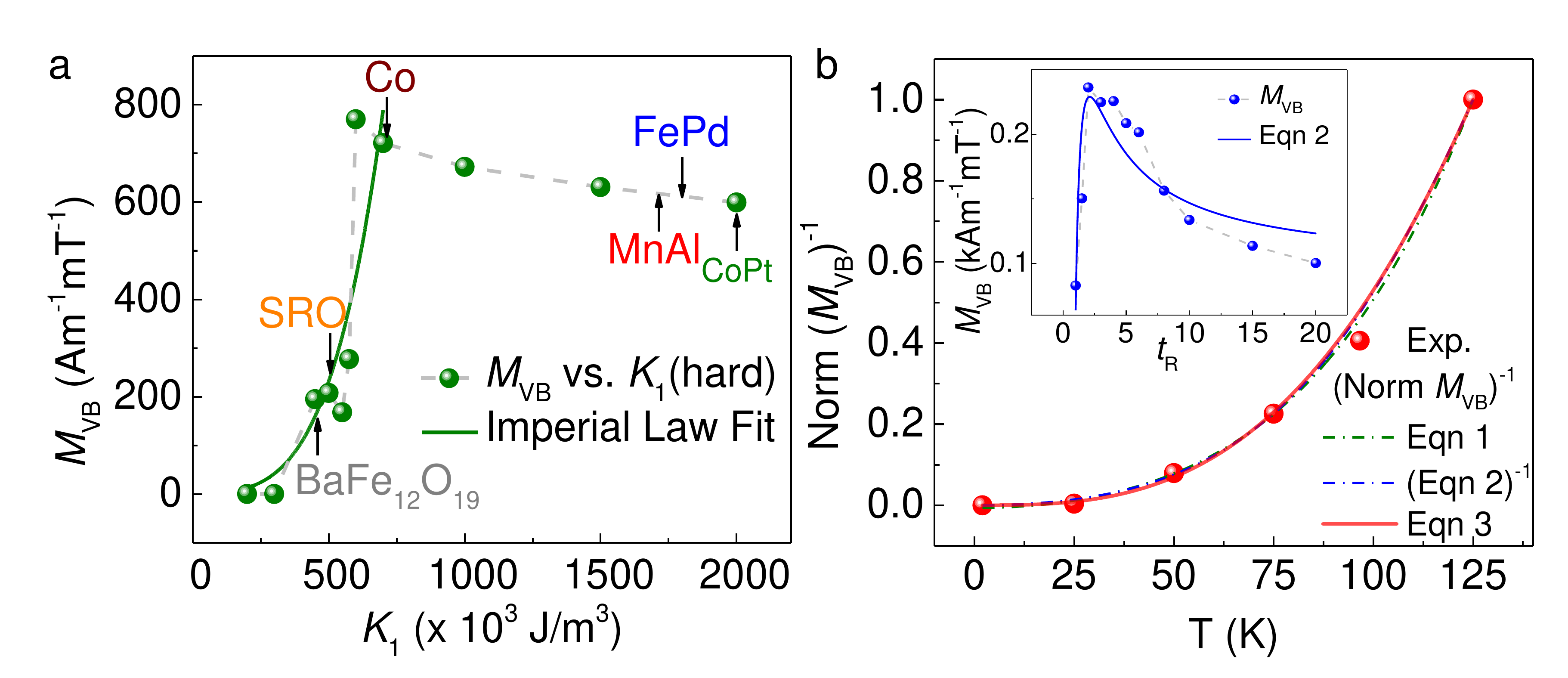}
	\vspace{-1em}
	\caption{(a) Dependence of $M_{\textrm{VB}}$ over $K_{\textrm{1}}$ fitted by the solid green curve (Imperial Law fit) from Eq. \eqref{eq1} for the data points (green spheres) obtained by simulations. Different hard materials are marked on the curve for their $K_{\textrm{1}}$ accordingly. (b) Dependence of $M_{\textrm{VB}}$ over $t_{\textrm{R}}$ is shown in inset. The solid blue line shows the best fitted curve by Eq. \eqref{eq2} (Spring Thickness fit) for the data points (blue spheres) achieved computationally. Correspondence of the reciprocal of  $T$ fit data with the reciprocal of $t$ fit (Spring Thickness fit) and $K_{\textrm{1}}$ fit (Imperial Law fit) independently and in combined form.}
	\vspace{-0.8em}
\end{figure}

Finally, we attempted to understand the relationship between temperature, anisotropy and thickness of the films which governs the VB. Here, using curve fittings and analysis of the experimental as well as theoretical results, we show a coherent dependence of the three parameters on each other. As we can clearly see from the Fig. 4(b), the reciprocal of the normalized experimental $M_{\textrm{VB}}$ curve for different temperatures exactly fits with the curves obtained from the reciprocal of the Spring Thickness fit from the thickness dependent simulated data, as well as independently with the Imperial Law fit of the simulated anisotropy data. Surprisingly, it has been seen that they not only fit the experimental ${M_{\textrm{VB}}}^{-1}$ \emph{vs.} $T$ data individually but in the combined form as well. This implies that the function governing the reciprocal of the shift gained from the temperature data is directly proportional to the product of the function of $M_{\textrm{VB}}$ for anisotropy (Eq. \eqref{eq1}) and the inverse function of $M_{\textrm{VB}}$ for thickness (Eq. \eqref{eq2}). We could express this perfectly fitted data in combined form by Eq. \eqref{eq3}. 
\begin{equation}
      M_{\textrm{VB}}^{-1} =  G' \frac{A'+B'{T^2}+C'{T^4}}{D'+E'/T+F'/T^2}      \label{eq3}
\end{equation}
where $A'-G'$ are the required fitting parameters (provided in SI Table S1). Hence, it can be inferred that experimental and theoretical results are absolutely fitting and complementing each other. Therefore, there is strong correlation between the parameters such as temperature, anisotropy and the thickness.

\section{Discussion}
To gain an insight into the mechanism of obtaining VB, we explain how a simple combination of hard and soft magnetic layers results into the novel VB without any complex phenomena being involved. The loop tracing field is of critical importance. SRO saturates at a very high field of $\sim$2.5 T\cite{ziese2010structural}, whereas Ni$_{80}$Fe$_{20}$ saturates at $\sim$20 mT\cite{zhou2000dependence} (Fig. 5(a)). Initially, an +FC (-FC) bias field aligns both the hard and the soft layer resulting into a net $M$ in the positive (negative) direction. Further, during field reversal for positive (negative) VB measurement, due to an insufficient amount of loop tracing field applied ($H<H_{\textrm{S}}$ (saturation field) of SRO), the effect of moment imprinted from the hard uncompensated spins of SRO in the non-collinear state cancels out from the moment of Ni$_{80}$Fe$_{20}$ because of the favourable anisotropy of the hard spins in the direction of FC bias. This decreases the negative (positive) moment and causes asymmetry in the loop. 
\vspace{-1em}
\begin{figure}[ht!]
 	\centering
 
 	\includegraphics[width=8.5cm]{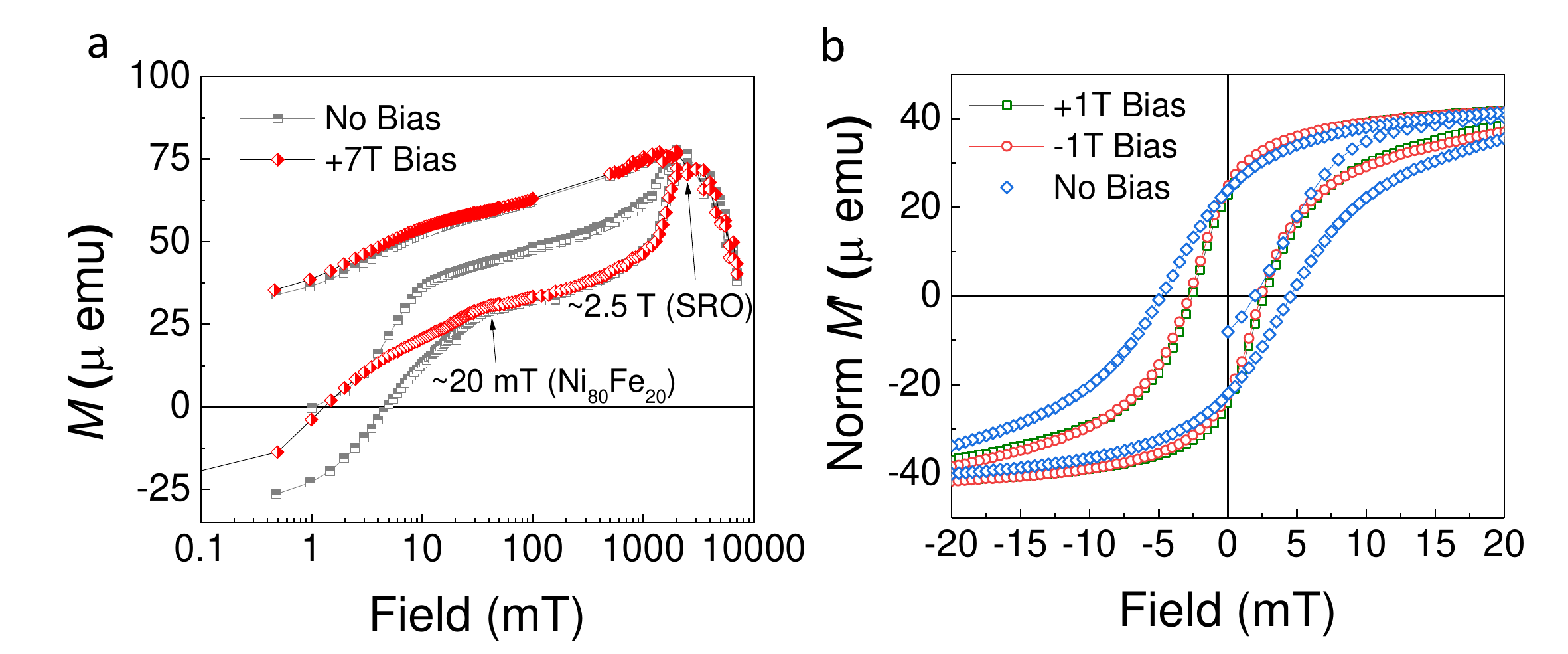}
 	\vspace{-1em}
	\caption{(a) Step-like hysteresis for the Ni$_{80}$Fe$_{20}$/SRO heterostructure for 7 T bias and no bias field. (b) VB subtracted $M'H$ curves illustrating a change in $H_{\textrm{C}}$ for ZFC-FC hysteresis loops but no horizontal EB. }
 \end{figure}
 
One another important observation is that the descending (ascending) branch of the positive (negative) biased $MH$ is exactly aligned over the descending (ascending) branch of the symmetric ZFC $MH$ near the coercive field region (where $M$ $\sim$ 0), clearly seen from Fig. 1 (experimental) and Fig. 2(b) (simulation). However, the corresponding opposite branches are not the parts of the symmetric loop. Hence, a significant decrease of $H_{\textrm{C}}$ (Fig. 1(b)) from $\sim$4.7 mT to $\sim$2.8 mT is measured experimentally for the vertically shifted FC hysteresis loops as compared to the non-shifted ZFC loop and observed in simulations as well (Fig. 2(b)). This observed decrease in $H_{\textrm{C}}$ can be understood as the pinning of the hard spins originating from FC in one direction (due to high anisotropy) under low field loop tracing range (±0.5 T). The frozen hard spins from FC favour the direction of external field when the loop is swept in the direction of FC bias. This now requires less field to achieve the required $M$, leading to low $H_{\textrm{C}}$ as compared to the case when loop tracing field direction opposes FC bias. Hence, the ascending (descending) branch for positive (negative) VB loop pulls away from the symmetric hysteresis loop.

Moreover, the observed horizontal shift ($\sim$1.7 mT) in Fig. 1 could be misunderstood as conventional EB at first instance. But interestingly, after removing the VB from the ±FC curves and plotting the $MH$, it is observed that there is no horizontal EB (Fig. 5(b)). Although, the decrease of $H_{\textrm{C}}$ for FC with respect to ZFC remains. This is similar to the remanence magnetization shift often observed in EB system due to curvature of $MH$ at $H=0$.  Hence, the VB does not originate from interfacial spin coupling like conventional EB rather due to the competition between the magnetic parameters. 
\begin{figure}[ht!]
	\centering
	\includegraphics[width=8.5cm]{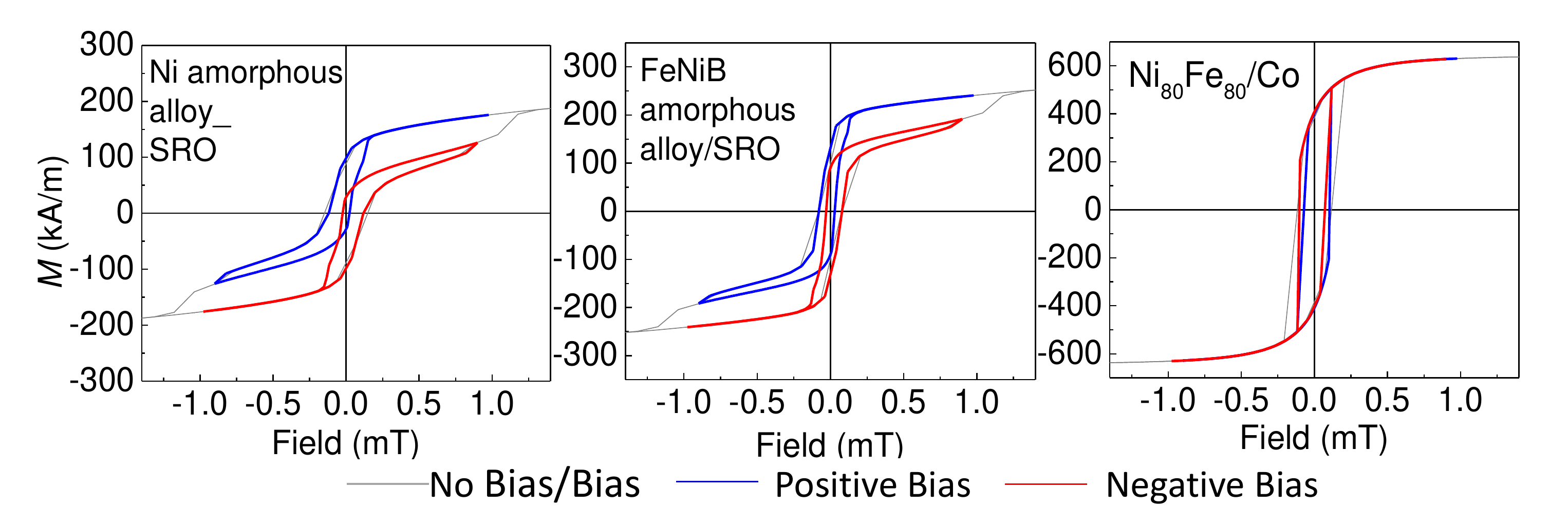}
 	\vspace{-1em}
 	\caption{$MH$ showing VB for different combination of soft and hard magnetic materials.}
\end{figure} 

Another interesting observation is that the positive (negative) $M_{\textrm{S}}$ of positive (negative) shifted curves coincides with the $+M_{\textrm{S}}$ ($-M_{\textrm{S}}$) of the fully saturated loop (refer Fig. 2) even at the low loop field tracing range. This is because the maximum saturation ($M_{\textrm{S}}$) for SRO is achievable for $T < T_{\textrm{C}}$ after applying a field (FC) at $T > T_{\textrm{C}}$ as SRO is PM at $T > T_{\textrm{C}}$ importantly. One can further use this result to exploit the applications where the maximum saturation needs to be achieved for VB without using very high fields similar to recent heat-assisted magnetic recording (HAMR) technology in modern hard disk drives\cite{wang2013hamr}. 

Nevertheless, volume fraction or $t_{\textrm{R}}$ plays an essential role in the whole picture. It was already shown, 2:1 is the $t_{\textrm{R}}$ for the maximum $M_{\textrm{VB}}$ in our case for the corresponding anisotropies of the two materials. The optimum thickness ratio could vary for different materials for different anisotropies. As we can see, for our case, $K_{\textrm{1}} \text{(hard layer)} > 600\; kJ/m^3$ leads to an ineffective change in $M_{\textrm{VB}}$ possibly due to the maximum compatibility ratio attained for the two layers at a certain point. A large increase of the hard anisotropy with the same $t_{\textrm{R}}$ may imbalance the ratio of the compensated spins to the uncompensated spins which may not change $M_{\textrm{VB}}$ much eventually.
\begin{table}
 \centering
 \caption{ Different materials combination for observing VB phenomenon.\\}
 \vspace{-1em}
    \begin{tabular}{ccccccc}
\hline
No. &	Material	& Type	& $M_{\textrm{S}}$ & $K_{\textrm{1}}$  & $A$ & Shift \\ 
  &   & & (kA/m) &  (kJ/m$^3$) &  (pJ/m) & \% \\
\hline

1 &  SRO & Hard & 350 & 500 & 1.8 & 54.7\\
 &  Ni-amorphous  & Soft & 493 & 0 & 8.5 & \\
  &  alloy &  & &  &  & \\
 \hline
2 &  SRO & Hard & 350 & 500 & 1.8 & 17.8\\
 &  FeNiB  & Soft & 836 & 0 & 8.07 & \\
  &   (40/40/20)- & &  &  &  &\\
  &  amorphous alloy & &  &  &  &\\
 \hline
3 &  Co & Hard & 1420 & 700 & 30.2 & 1.04\\
 &  Ni$_{80}$Fe$_{20}$ & Soft & 600 & 0 & 13 & \\
 \hline
\end{tabular}
\label{tab:table3}
 \vspace{-1em}
\end{table}
\vspace{0.5em}

To reproduce the proof of concept for different materials from our proposed model, we have also simulated for different combination of materials (e.g., Co/Ni$_{80}$Fe$_{20}$, SRO/Ni-amorphous alloy, etc.) with their individual characteristic parameters like $K_{\textrm{1}}$, $M_{\textrm{S}}$ and $A$ and found similar VB behaviour (refer to Table 1 and Fig. 6). It is clearly visible that Ni$_{80}$Fe$_{20}$/Co has come up with the minimum shift. This will allow us to choose the right material combination in different cases according to the application point of view. 

\section{Conclusion}
We have demonstrated and modelled a simple bilayer thin film heterostructure of a soft and a hard layer to obtain a large VB shift. By experiment and detailed micromagnetic simulations, we have observed that the important parameters for achieving the large VB are the respective anisotropies and thickness ratio of the two layers. An optimum amount of loop tracing field range below $T_{\textrm{C}}$ of the hard material is also essential to achieve the vertical shift. We have developed a generalized model by fitting the data incorporating both experimental and simulation results, for the first time to demonstrate achievable VB by different bilayer material combinations. This VB system is much simpler, easy to reproduce, tuneable and controllable compared to alternative systems, including complex FM/AFM spin glasses, or doped magnetic systems, or systems containing vacancies or topological defects. The proposed model can also be applicable to the other VB systems with quantified spin glass, pinned/uncompensated spins or coupled interfaces. The work is an important step forward to a comprehensive understanding of tuneable VB and therefore has important implications in future spintronic devices. 

\section*{Acknowledgement}
MB and TM acknowledge IISER Thiruvananthapuram, India for funding and support for HPC time at the Padmanabha cluster, IISER Thiruvananthapuram, India. We are also funded from the Leverhulme Trust grant $\#$ RPG-2015-017, the EPSRC grants EP/N004272/1, and EP/M000524/1, the Royal Academy of Engineering Chair in Emerging Technologies Grant CiET1819$\backslash$24, the EU grant H2020-MSCA-IF-2016 (745886)-MuStMAM, and the Isaac Newton Trust (RG96474).

\newpage
\onecolumngrid
\Huge Additional Information \par
\vspace{1em}
\normalsize
This additional Information contains the details of the micromagnetic simulations performed to effectuate the phenomena of novel vertical bias. \\

\textbf{A. Simulation protocols}: Initially, a 10 nm layer of Ni$_{80}$Fe$_{20}$ film and 20 nm of SRO film were simulated individually to estimate the correctness of the $H_{\textrm{C}}$ and the $M_{\textrm{S}}$ of the two films (shown in Fig. 7). $H_{\textrm{C}}$ for Ni$_{80}$Fe$_{20}$ and SRO is noted to be 8 mT and 1.48 T, respectively which are close to the experimental values\cite{huang2015augmentation,zhou2000dependence,bryan2004experimental}. Further, a bilayer film of 10 nm Ni$_{80}$Fe$_{20}$ (top) and 20 nm SRO (bottom) was simulated for obtaining the VB. Later on, various parameters including $t_{\textrm{R}}$ of the two films, $K_{\textrm{1}}$, etc., were varied to estimate their effect on $M_{\textrm{VB}}$. Fig. 9 shows the trend of change of $H_{\textrm{C}}$ with respect to $t_{\textrm{R}}$ and $K_{\textrm{1}}$.

 \begin{figure}[ht!]
	\centering
	\includegraphics[width=12.6cm]{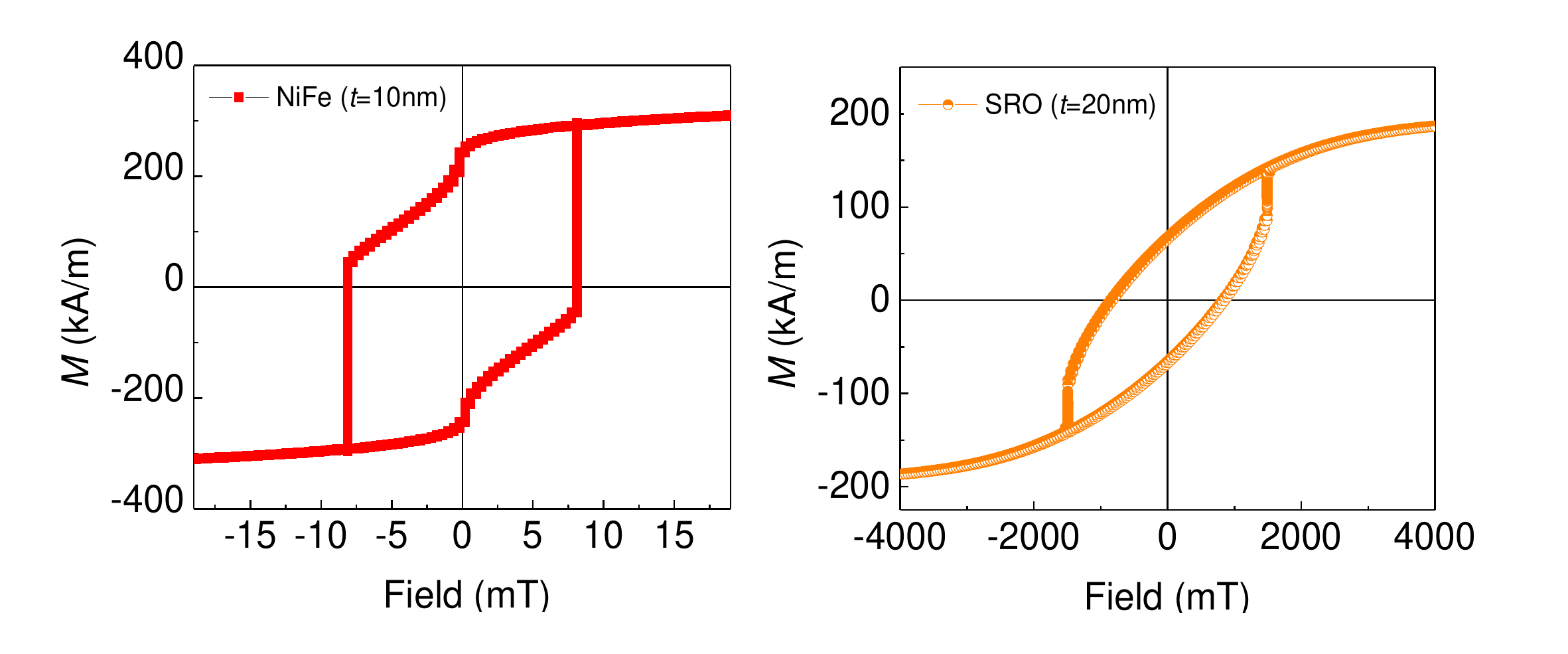}
 	\vspace{-1em}
 	\caption{Magnetic hysteresis loop for Ni$_{80}$Fe$_{20}$ (10 nm) : $H_{\textrm{C}}$ = 8 mT. Magnetic hysteresis loop for SRO (20 nm) : $H_{\textrm{C}}$ = 1.48 T.}
 	\vspace{-0.7em}
 \end{figure}
 
\begin{figure}[ht!]
	\centering
	\includegraphics[width=9cm]{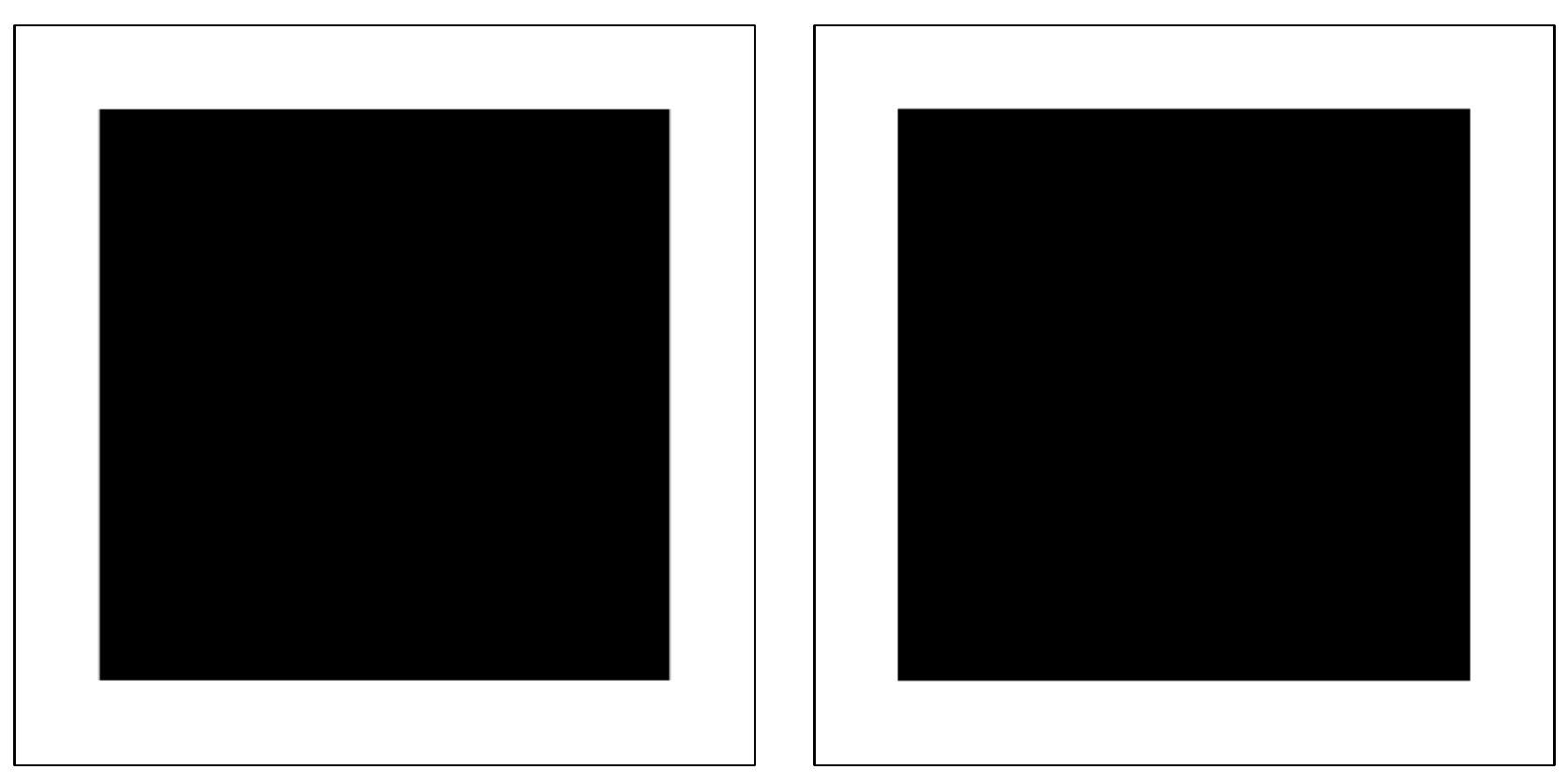}
 	\vspace{-1em}
 	\caption{Bitmaps used in simulations for the two films: Ni$_{80}$Fe$_{20}$ (320 nm x 320 nm x 10 nm) and SRO (320 nm x 320 nm x 20 nm).}
 	\vspace{-0.7em}
\end{figure}

\begin{figure}[ht!]
	\centering
	\includegraphics[width=12.6cm]{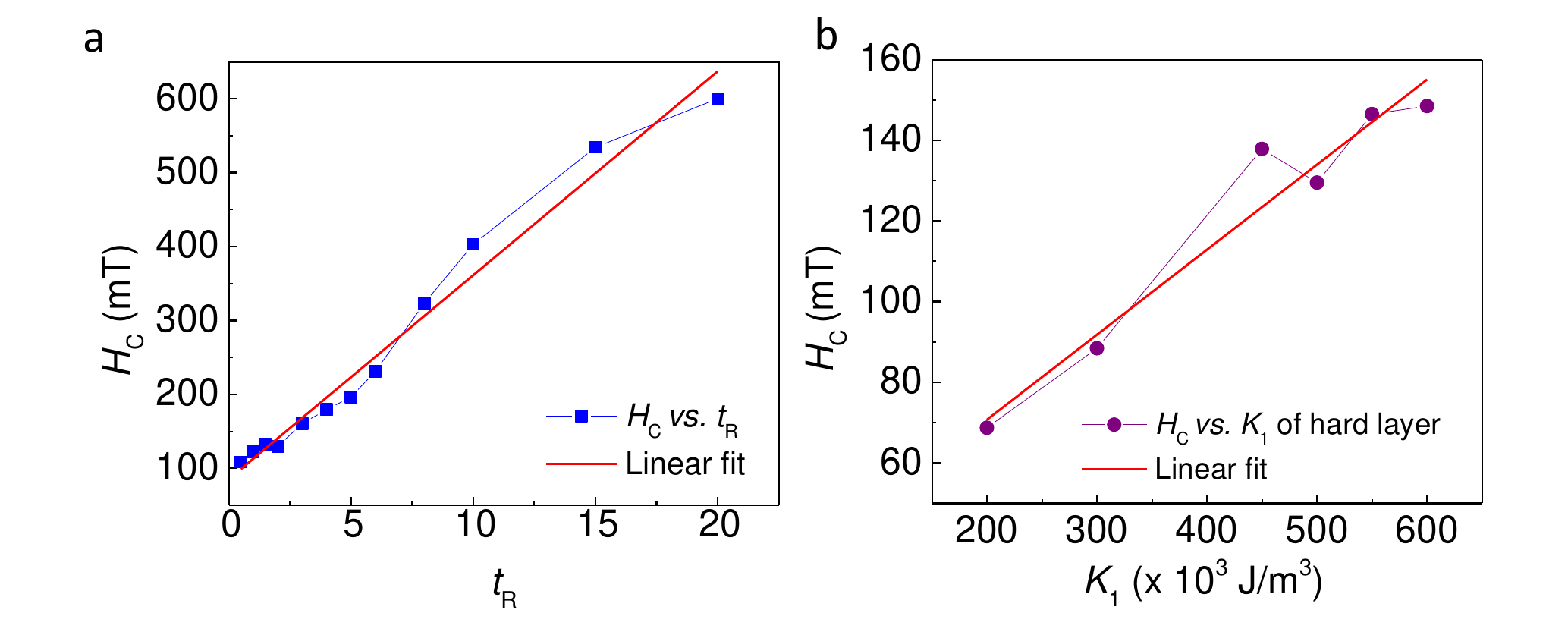}
 	\vspace{-1em}
 	\caption{Variation of $H_{\textrm{C}}$ with $t_{\textrm{R}}$ and $K_{\textrm{1}}$ of hard layer.}
\end{figure}
\newpage
\textbf{B.	Thickness variation}:
Fig. 10 shows few plots representing varied $t_{\textrm{R}}$ (other parameters are constant) of the two magnetic layers, i.e., SRO and Ni$_{80}$Fe$_{20}$ to achieve the maximum $M_{\textrm{VB}}$. The black loop is the fully saturated loop achieved with loop field range ±7 T whereas red (blue) corresponds to the loops with field sweep range ±0.9 T whose unsaturated spins have gained the VB. It has been found that the maximum has been obtained for the $t_{\textrm{R}}$ 2:1 for SRO (hard) and Ni$_{80}$Fe$_{20}$ (soft) layer respectively (highlighted in red text) which is similar to the ratio taken experimentally (2 nm for Ni$_{80}$Fe$_{20}$ and 5 nm for SRO) too. As we go for very high ratios, the overall shift is decreasing. The increasing effect of the hard layer for higher $t$ of SRO decreases the effectiveness of the soft layer. Such an imbalance reduces the $M_{\textrm{VB}}$. \\

\begin{figure}[ht!]
	\centering
	\includegraphics[width=13cm]{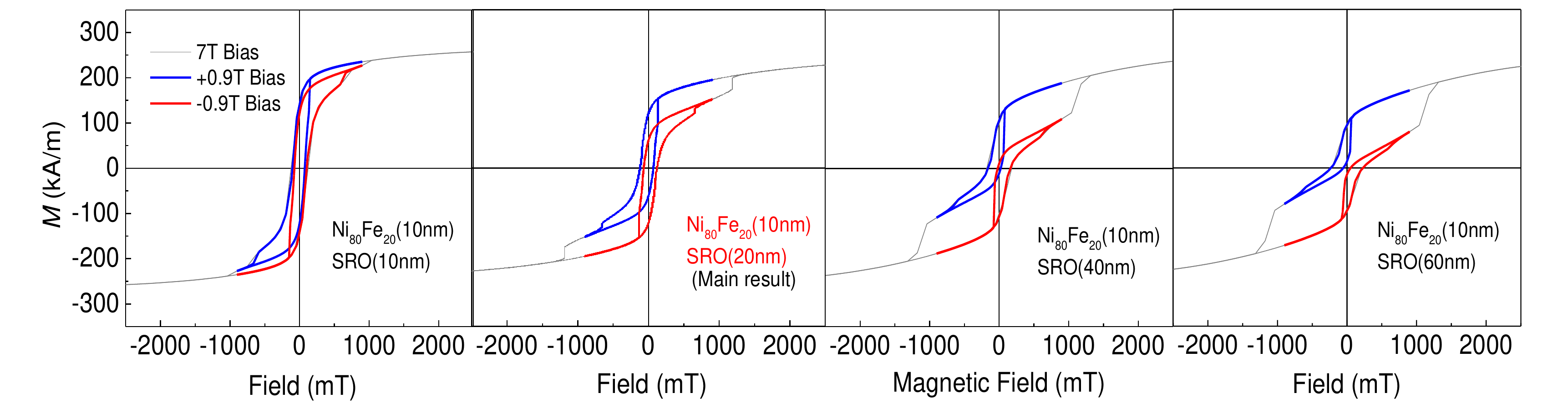}
 	\vspace{-1em}
 	\caption{$MH$ showing VB for the bilayer system of  Ni$_{80}$Fe$_{20}$/SRO for $t_{\textrm{R}}$ ranging from 1:1 to 6:1. Ratio 2:1 (text highlighted in red) is approximately same as that of experimental ratio of the two layers.}
 	
\end{figure}

\textbf{C.	Anisotropy constant variation}:
$K_{\textrm{1}}$ of the hard layer was varied while keeping the other parameters constant by the help of simulations. Some examples are shown in Fig. 11. Almost negligible vertical shift is seen for the anisotropy of hard layer as 200 kJ/m$^3$ (shown in the first graph of Fig. 11). Anisotropy constant $K_{\textrm{1}}$=500 kJ/m$^3$ (highlighted in red) matches most with the experimental results. Increasing $K_{\textrm{1}}$ further doesn't change $M_{\textrm{VB}}$ much, and hence is not considered. \\

\begin{figure}[ht!]
	\centering
	\includegraphics[width=13cm]{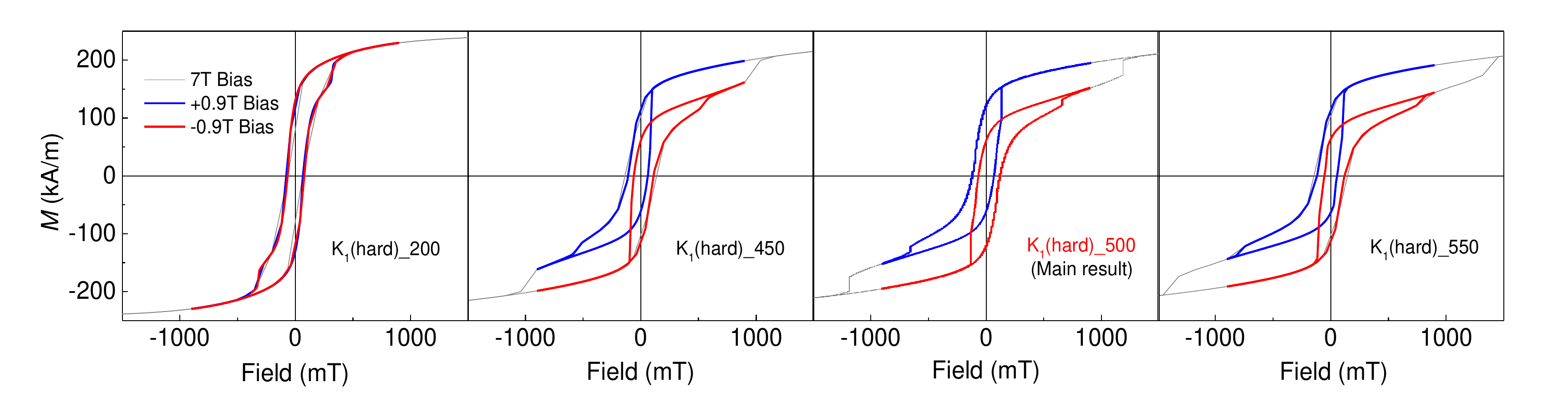}
 	\vspace{-1em}
 	\caption{$MH$ showing VB for the bilayer system of  Ni$_{80}$Fe$_{20}$/SRO for different anisotropy constants of the hard layer. Anisotropy of the soft layer was kept constant as 0 J/m$^3$. Anisotropy constant $K_{\textrm{1}}$ = 500 kJ/m$^3$ (text highlighted in red) is the anisotropy of SRO serving as our main result.}
\end{figure}

\textbf{D.	Anisotropy direction variation}:
According to the Stoner- Wohlfarth model\cite{agrawal2018magnetic}, the overall energy of a system is expressed by Eq. \eqref{4} as:
\begin{equation}
                        E= K_{\textrm{U}}sin(\phi-\theta)^2-\mu MHcos(\phi)   \label{4}
\end{equation}                                   
where $K_{\textrm{U}}$ is the uniaxial anisotropy constant, $M$ is the magnetization, $H$ is the external magnetic field, $\theta$ and $\phi$  are the angles of magnetic field with anisotropy and magnetization, respectively (refer to Fig. 12.). 
\begin{figure}[ht!]
	\centering
	\includegraphics[width=5cm]{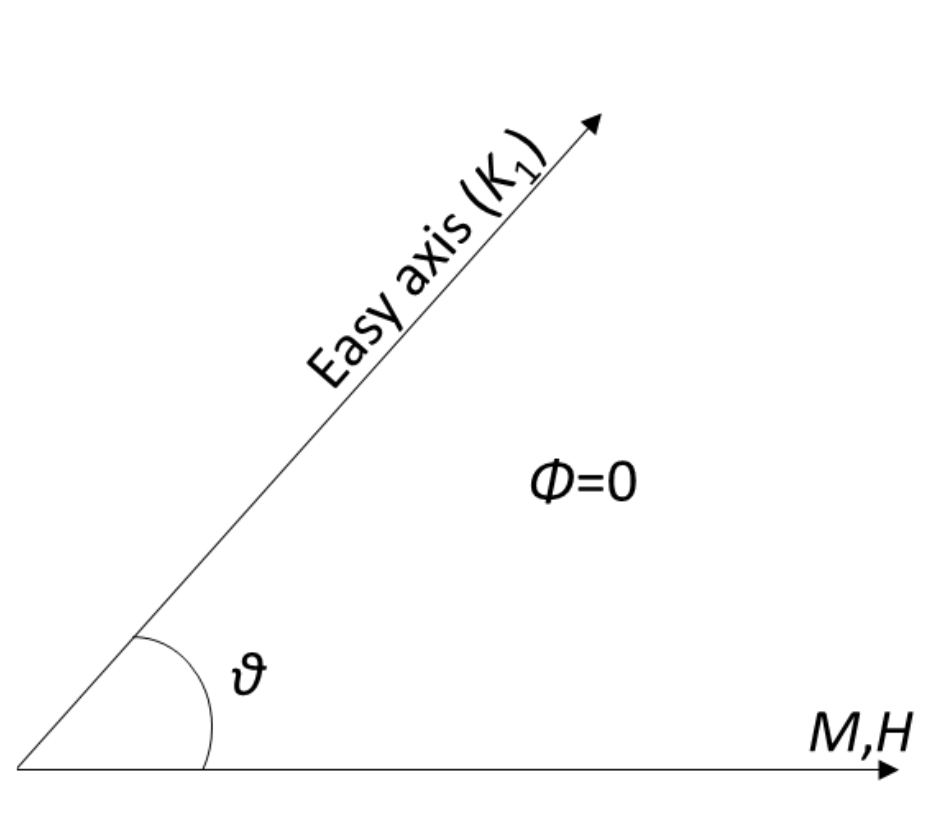}
 	\vspace{-1em}
 	\caption{Magnetization and applied external magnetic field is in the same direction implying $\phi=0$. Anisotropy easy axis is at an angle $\theta$ with the magnetization/magnetic field.}
\end{figure}
The first and second term in Eq. \eqref{4} refer to the anisotropy energy ($E_{\textrm{A}}$) and Zeeman energy ($E_{\textrm{Z}}$)  and both the energies compete with each other. Anisotropy energy is crucially defined by the direction of the easy axis. If the anisotropy easy axis is in the direction of applied magnetic field, then it is easy for the spin moments to align in the common direction which in effect will give a constant $M$ on applying a very small amount of $H$, giving a square-like hysteresis loop. This is because both $E_{\textrm{Z}}$  and $E_{\textrm{A}}$ are favouring each other parallelly. However, when the magnetic field is applied along the hard axis, both the energies start competing with each other. It can be further understood from the Eq. S1 depicting an angular dependence between the two energies.
\begin{figure}[ht!]
	\centering
	\includegraphics[width=13cm]{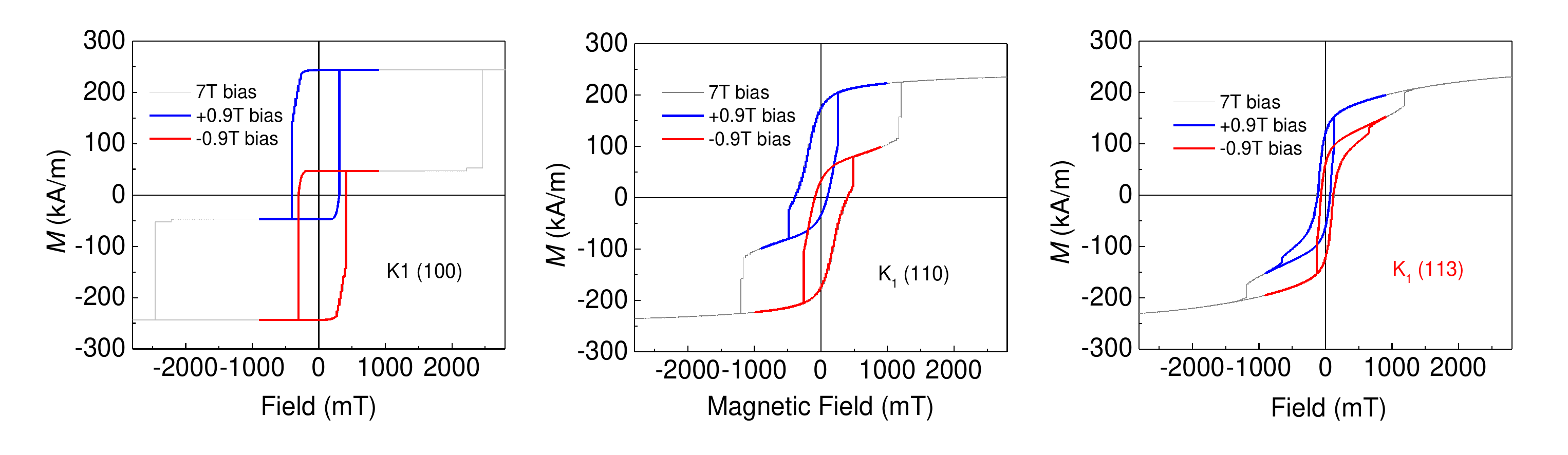}
 	\vspace{-1em}
 	\caption{$MH$ showing $M_{\textrm{VB}}$ for the bilayer system of  Ni$_{80}$Fe$_{20}$/SRO for different directions of anisotropy of the hard layer i.e., (100), (110) and (113). Anisotropy direction (113) (text highlighted in red) is the anisotropy direction of SRO taken for the main result.}
\end{figure}

For our case, the initial $M$ axis was kept same as that of the external $H$ direction, i.e., ((100) ; X axis). We have tried changing the anisotropy direction such as (100), (110) and (113) (depicted in Fig. 13). The VB shift is observed for all three directions. However, direction (100) shows the constant square-like $MH$ whereas direction (113) gives the most comparable result to the experimental outcomes, therefore we have chosen (113) for anisotropy direction as our most significant result. \\

\begin{table}[h]
 \centering
 \caption{ Fitting parameters of modelling for Eq. 1, 2 and 3 respectively in the main text. }
 \vspace{1em}
    \begin{tabular}{cccc}
\hline
Fitting parameters &	Equation 1	& Equation 2	& Equation 3\\
\hline
$A/A'$ &	0 &	- &	-0.00642\\
$B/B'$ &	2.10736 x 10$^{-4}$  &	- &	229.36768 \\
$C/C'$ & 2.85514  x 10$^{-9}$ &	- &	-0.00131 \\
$D/D'$ &	-	 & 96.8043 &	1.96123 \\
$E/E'$ &	- &	560.341 &	2.23799 x 10$^7$\\
$F/F'$ &	- &	-593.348 &	57011 \\
$G'$ &	-&	-	& 0.05486 \\
\hline
\end{tabular}

\label{tab:table2}
\end{table}
\newpage
\bibliographystyle{unsrtnat}
\bibliography{bibliography}
\end{document}